\documentclass{article}
\usepackage{spconf,amsmath,graphicx,hyperref}
\usepackage{booktabs}
\usepackage[table]{xcolor}
\usepackage{amssymb}
\usepackage{bm}
\usepackage{graphicx}
\usepackage{svg}
\usepackage{algorithm}      
\usepackage{algpseudocode}  

\newif\ifarxiv
\arxivtrue      

\title{Semantic Sound Separation with Latent Mixture Invariant Training}
\name{Dimitrios Bralios$^{1*}$\thanks{*Work done while at Google DeepMind.}, Henry Li$^{2}$, Robin Scheibler$^{2}$, Arnaud Doucet$^{2}$, John R. Hershey$^{2}$}
\address{$^{1}$University of Illinois Urbana-Champaign \qquad $^{2}$Google DeepMind}

\begin{document}
\ninept
\maketitle
%
\begin{abstract}
Unsupervised audio source separation via Mixture Invariant Training (MixIT) enables training on unlabeled acoustic mixtures, but lacks semantic grounding and is vulnerable to over-separation. Conversely, text-conditional source separation achieves strong semantic control but requires supervised single-source audio-text pairs and manual text prompts at inference. 
In this work, we propose \textit{latent MixIT}, a prompt-free, mixture-supervised framework for learning source-specific semantic embeddings directly from audio mixtures.
By extending the standard waveform MixIT objective to a pre-trained audio embedding space, our method learns latent source representations without requiring isolated source signals. 
When combined with an audio-domain separation objective, latent MixIT reduces over-separation and improves separation across multiple architectures. We further introduce the FUSS Retrieval Benchmark for source aware zero-shot audio-text retrieval and show that the learned source embeddings support strong single-pass performance, closing much of the gap to oracle embeddings from ground-truth sources and clearly surpassing unsupervised embeddings on low-energy sources.
\end{abstract}

\begin{keywords}
audio source separation, source embeddings, MixIT, audio-text retrieval, representation learning
\end{keywords}

\section{Introduction}
\label{sec:intro}
\renewcommand{\paragraph}[1]{\noindent\textbf{#1}\quad}

A hallmark of human auditory scene analysis is the ability to parse complex acoustic environments into distinct sound sources~\cite{bregman1990auditory}. In recent years, deep learning has driven rapid progress in computational audio understanding. State-of-the-art models classify audio events over predefined label spaces~\cite{gong2021ast, gemmeke2017audio} or map audio to text embeddings via contrastive objectives, e.g. CLAP~\cite{wu2023large, elizalde2023clap}. However, these approaches produce clip-level representations that compress the entire acoustic mixture into a single global descriptor. When multiple sound events co-occur, their semantic properties are entangled rather than resolved into source-specific entities.

Conversely, single-channel sound separation aims to disentangle individual source signals directly from acoustic mixtures~\cite{luo2018tasnet, luo2019conv}. Supervised methods such as permutation invariant training (PIT)~\cite{yu2017permutation} achieve strong separation, but rely on synthetic mixtures formed from isolated source signals. This synthetic training regime introduces domain shift on real-world recordings and focuses purely on signal-level waveform extraction rather than semantic representation learning. Hybrid paradigms—such as conditioning separation on inferred speaker identities~\cite{zeghidour2021wavesplit} or language queries~\cite{liu2022separate, liu2024separate}—partially bridge this gap, but still require supervised annotations, isolated targets, or explicit text prompts during inference. While mixture invariant training (MixIT)~\cite{wisdom2020unsupervised} removed the requirement for isolated reference signals by training models to separate mixtures of mixtures (MoMs) into latent waveform sources. 

Recent work has explored multi-source representation learning through contrastive learning with separated musical stems~\cite{garoufis2023multi}, slot-based reconstruction of frozen encoder features~\cite{sridhar2025compositional}, and a representation space MixIT formulation in SSLAM~\cite{alex2025sslam}. 
Most closely related, SSLAM explored predicting multiple representations from mixed audio and the use of a MixIT objective for the task of audio classification, however, in preliminary experiments, this formulation underperformed their primary method.

In this paper, we investigate whether the MixIT principle can be generalized from low-level signal reconstruction to high-level representation learning. We propose latent MixIT, an unsupervised framework that operates on pre-trained semantic embedding spaces, such as CLAP space~\cite{wu2023large, elizalde2023clap}. Analogous to waveform MixIT, latent MixIT forms a mixture-of-mixtures in waveform space and trains a model to decompose it into a set of source-specific latent slots that remix to reconstruct the constituent mixture embeddings, eliminating the need for isolated ground-truth sources.

Our primary contributions are:
\begin{itemize}
    \item Latent Mixture Invariant Training: We formulate the mixture-invariant objective over semantic embeddings, enabling unsupervised decomposition of multi-source audio mixtures into source-specific representations.
    \item Dual Stage Latent-to-Waveform Separation: We integrate latent MixIT with conventional audio-domain MixIT into a unified two-stage pipeline, demonstrating that semantic slot conditioning directly improves waveform separation quality.
    \item FUSS Retrieval Benchmark: We introduce a zero-shot audio-text retrieval benchmark on FUSS~\cite{wisdom2021s}, where our learned source slots attain the highest unsupervised R@10 (73.2\%) and substantially outperform separated waveforms on low-energy sources. 
\end{itemize}

\section{Background}
\label{sec:background}

\begin{figure*}[t]
  \centering
  
  \vspace{-1.6cm}
  \includegraphics[width=0.8\textwidth]{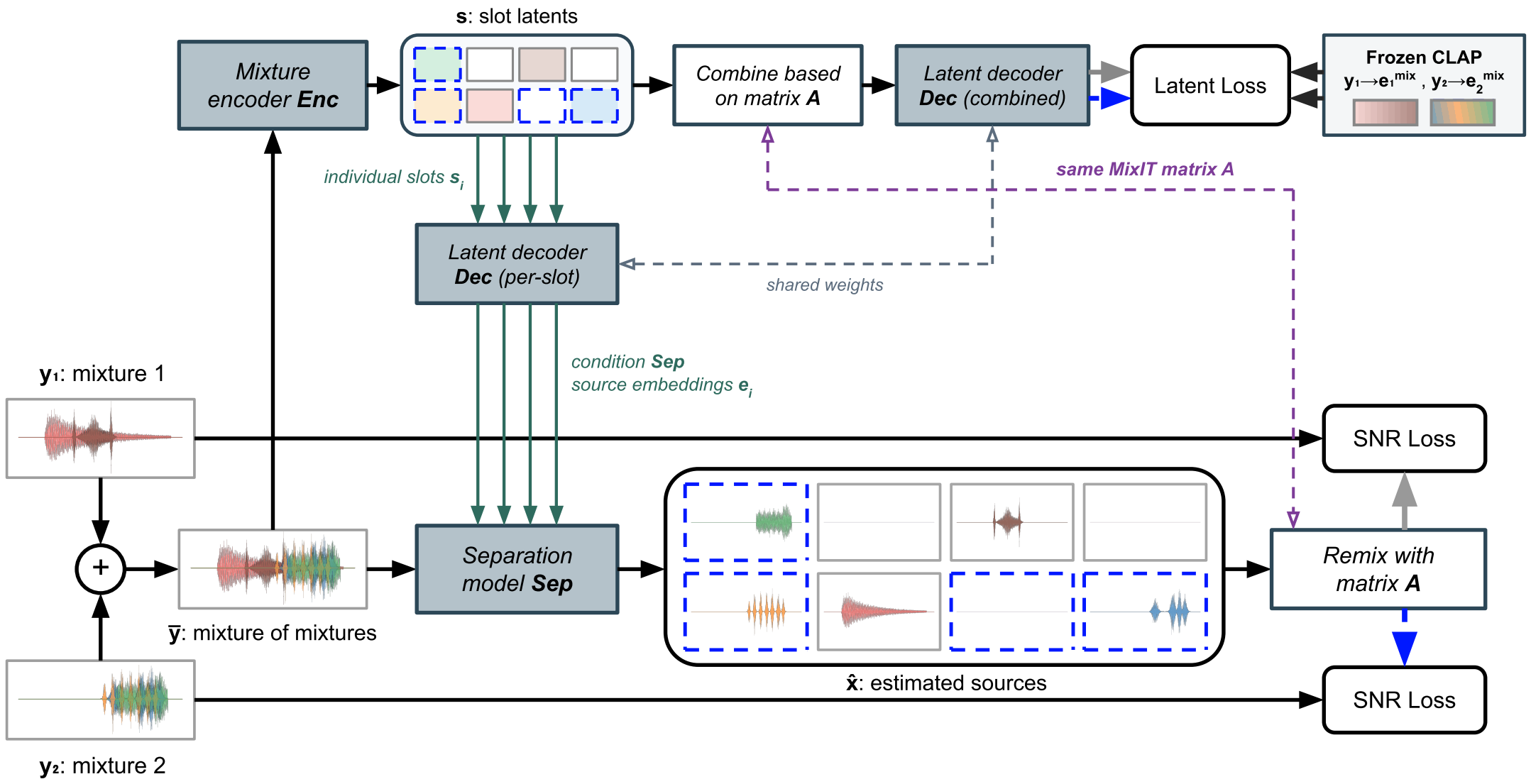}
  \vspace{-0.2cm}
  \caption{\textbf{Joint latent-and-waveform MixIT training pipeline.}
  From the mixture of mixtures $\bar{\bm{y}} = \bm{y}_1 + \bm{y}_2$, the \emph{Mixture encoder} produces $N$ slot latents $\bm{s}_i$ that pass through a shared \emph{Latent decoder} along two paths:
  \textbf{(a)}~individually to condition the waveform separator $\mathrm{Sep}$ when estimating sources $\hat{\bm{x}}$, and
  \textbf{(b)}~combined via the MixIT assignment matrix $\mathbf{A} \in \{0,1\}^{2 \times N}$ to match frozen CLAP targets $(\bm{e}^\mathrm{mix}_1, \bm{e}^\mathrm{mix}_2)$ of $(\bm{y}_1, \bm{y}_2)$ under the \emph{Latent Loss}.
  The same assignment matrix $\mathbf{A}$ (grey solid vs.\ blue dashed slots) jointly governs latent combination and waveform remixing ($\mathbf{A}\hat{\bm{x}}$) for the \emph{SNR Loss}.}
  \label{fig:model_diagram}
  \vspace{-0.3cm}
\end{figure*}

In this section, we first review MixIT~\cite{wisdom2020unsupervised} before introducing our extension of it to source-based latent representations in the following Section.


MixIT enables source separation training without access to isolated source signals, allowing observed mixtures to be used as supervision instead. Given two mixtures $\bm{y}_1$ and $\bm{y}_2$, we construct a mixture-of-mixtures $\bar{\bm{y}} = \bm{y}_1 + \bm{y}_2$, which is fed into a separator $\mathrm{Sep}: \mathbb{R}^T\!\to\!(\mathbb{R}^T)^N$, producing a collection of $N$ estimated sources,
\begin{equation}
\{\hat{\bm{x}}_i\} = \mathrm{Sep}(\bar{\bm{y}}),
\end{equation}
where we drop the explicit bounds $i=1, \dots, N$ for brevity. Since the mapping between the $N$ estimated sources and the two original mixtures is unknown, MixIT considers all possible assignments each represented by a binary assignment matrix $\bm{A} \in \{0,1\}^{2 \times N}$, where $\sum_{m=1}^{2} A_{m,i}=1$ for all $i$, and $\sum_{i=1}^N A_{m,i} \geq 1$ for all $m$. Each assignment partitions the estimated sources into two groups, whose sums are used to reconstruct the corresponding original mixtures.

Finally, the MixIT training objective selects the assignment that best reconstructs the two original mixtures, given a signal-level reconstruction loss $\mathcal{L}$,
\begin{equation}
\mathcal{L}_{\mathrm{MixIT}}
=
\min_{\bm{A}}
\sum_{m=1}^{2}
\mathcal{L}
\left(
\bm{y}_m,
\sum_{i=1}^{N} A_{m,i}\hat{\bm{x}}_i
\right),
\label{eq:audio_mixit}
\end{equation}
Assuming the sources that make up the original mixtures are independent, this training scheme allows the separator to learn a source decomposition using mixtures as supervision.

\section{Proposed Method}
\label{sec:method}

\subsection{Latent Space MixIT}

Given an audio mixture $\bm{y} \in \mathbb{R}^T$ comprising a collection of up to $N$ sources $\{\bm{x}_i \}$, with $\bm{x}_i \in \mathbb{R}^T$, our goal is to extract $N$ source-based embeddings $\{\bm{e}_i\}$, where each $\bm{e}_i \in \mathbb{R}^D$ is intended to capture the semantic content of a potential source present in the mixture.

To achieve this under the same MixIT training setup, we extend the formulation to the latent space, where an encoder model $\mathrm{Enc}\colon \mathbb{R}^T \!\to\!(\mathbb{R}^C)^N$ produces a collection of $N$ source-based embeddings instead of source waveforms.

During training, given the mixture-of-mixtures $\bar{\bm{y}}$, the model produces a collection of $N$ slot embeddings
\begin{equation}
 \{\bm{s}_i\} = \mathrm{Enc}(\bar{\bm{y}}).
\end{equation}
Each latent slot $\bm{s}_i \in \mathbb{R}^C$ is intended to represent one potential source in the input mixture.

In the case of MixIT the original mixtures are directly used as targets. In the latent setting, we instead extract embedding targets by using a frozen pre-trained embedding model $\mathrm{Emb}$ on $\bm{y}_1$ and $\bm{y}_2$.
\begin{equation}
\bm{e}^\mathrm{mix}_1 = \mathrm{Emb}(\bm{y}_1), \quad \bm{e}^\mathrm{mix}_2 = \mathrm{Emb}(\bm{y}_2).
\end{equation}

As in MixIT, the correspondence between the predicted latent slots and the constituent mixtures is unknown. We therefore use an assignment $\bm{A}$ to group the $N$ latent slots according to the mixture they are associated with. 

However, unlike waveforms, embeddings are not additive. Therefore, the slots assigned to a mixture cannot simply be summed and directly compared with its target embedding. Instead, we introduce a decoder $\mathrm{Dec}\colon (\mathbb{R}^C)^+\!\!\rightarrow\!\mathbb{R}^D$ that combines a variable number of source encodings into a single mixture embedding,
\begin{equation}
\hat{\bm{e}}^\mathrm{mix}_m =
\mathrm{Dec}\left(
\{\bm{s}_i \mid A_{m,i}=1\}
\right).
\end{equation}
The resulting embedding $\hat{\bm{e}}^\mathrm{mix}_m$ can then be compared with the corresponding mixture embedding target $\bm{e}^\mathrm{mix}_m$.

The latent MixIT objective selects the assignment that best matches the latent representations associated with the two constituent mixtures,
\begin{equation}
\mathcal{L}_{\mathrm{latent}}
=
\min_{\bm{A}}
\sum_{m=1}^{2}
\mathcal{L}_{\mathrm{emb}}
\left(
\bm{e}^\mathrm{mix}_m,
\hat{\bm{e}}^\mathrm{mix}_m
\right),
\label{eq:latent_mixit}
\end{equation}
where $\mathcal{L}_{\mathrm{emb}}$ denotes our embedding-level objective. In contrast to standard MixIT, which constrains the model through waveform reconstruction, latent MixIT directly encourages the learned slots to form source-specific semantic representations.

The resulting latent slots can be mapped into the pre-trained embedding space using the same decoder $\mathrm{Dec}$, applied independently to each slot, resulting in $\bm{e}_i = \mathrm{Dec}(\bm{s}_i)$. This allows the source embeddings to be used directly for downstream semantic tasks.

\subsection{Two-Stage Separation}

We next combine the latent-space formulation with audio-domain separation in a two-stage cascaded system. The first stage decomposes the input mixture into a set of source embeddings, which are then fed into an equivariant separator that produces source estimates. This allows the model to combine supervision from the audio and latent spaces. More concretely, a forward pass through the model is defined as
\begin{align}
 \{\bm{s}_i\}  &= \mathrm{Enc}(\bar{\bm{y}}),\\
\{\hat{\bm{x}}_i\} &= \mathrm{Sep}\left(\bar{\bm{y}}, \{\mathrm{Dec}(\bm{s}_i)\}\right).
\end{align}The training objective is defined as
\begin{equation}
\mathcal{L}_{\mathrm{two-stage}}
=
\min_{\bm{A}}
\left[
\mathcal{L}_{\mathrm{audio}}(\bm{A})
+
\lambda \mathcal{L}_{\mathrm{latent}}(\bm{A})
\right].
\end{equation}
Here, $\mathcal{L}_{\mathrm{audio}}(\bm{A})$ and $\mathcal{L}_{\mathrm{latent}}(\bm{A})$ denote the assignment-conditioned objectives inside the minimizations of Eqs.~\ref{eq:audio_mixit} and~\ref{eq:latent_mixit}, respectively. In the two-stage formulation, both terms are evaluated using the same assignment $\bm{A}$, which is selected jointly by the combined objective.

\section{Experimental Setup}
\label{sec:exp}

\subsection{Embedding Model}

Our embedding model $\mathrm{Emb}$ is a pre-trained CLAP model~\cite{wu2023large}, utilizing an HTS-AT~\cite{chen2022hts} audio encoder and a RoBERTa-based~\cite{liu2019roberta} text encoder pre-trained on the LAION-Audio-630k dataset. Both audio and text modalities are projected into a $D=512$-dimensional shared embedding space. $\mathrm{Emb}$ is kept strictly frozen throughout training and is used to extract fixed reference target embeddings for the mixtures $\bm{y}_1$ and $\bm{y}_2$.

\subsection{Slot Encoder and Decoder Architecture} 

Our slot encoder $\mathrm{Enc}$ follows a TasNet-style architecture~\cite{luo2018tasnet, kavalerov2019universal}. 
A log-mel input with 64 bins is projected to 256 dimensions and processed by a TDCN++~\cite{kavalerov2019universal} backbone with 4 superblocks of 8 dilated residual blocks each. 
The resulting features are projected into $N=4$ slot-specific sequences and time-pooled using attention, producing one $C=256$-dimensional representation per slot.

Our decoder $\mathrm{Dec}$ maps slot subsets to the pre-trained CLAP space. For each MixIT assignment, the selected slots and a learnable $\mathrm{CLS}$ token are processed by a single-layer 4-head Transformer ($256$ dims, FFN $=1024$), the $\mathrm{CLS}$ output is projected to 512 dimensions and unit-normalized. During training, $\mathrm{Dec}$ is evaluated over all valid MixIT assignments. At inference, it is only applied independently to each slot to obtain its corresponding source embedding.

\subsection{Separator Architecture} 

In experiments, we use conditional and permutation equivariant separators, maintaining a consistent correspondence between conditioning vectors and source estimates. 
All architectures predict source-specific real masks over the mixture magnitude.
We consider two architectures that differ in cross-slot interaction. 
The first processes each slot independently, using a shared separation core following a TasNet-style TDCN++~\cite{kavalerov2019universal} architecture, with cross-slot interaction only at mask estimation.
The second allows explicit inter-slot interaction through self-attention, with a Locoformer-based~\cite{saijo2024tf} core augmented with an additional slot dimension. 
No positional encodings are used over the slot dimension.

In order to match model capacity across comparisons, most experiments use the full cascaded separation model, including the slot encoder and decoder, even when $\lambda=0$ i.e., pure waveform MixIT. 
The PIT baseline instead replaces the encoder and decoder with learned conditioning vectors while retaining the same separator.

\subsection{Training Setup}

Models are trained for $250\text{k}$ steps with a batch size of 64 using Adam~\cite{kingma2014adam}, a learning rate of $5\times10^{-4}$, and 500-step linear warmup. 
We train on $\approx300\text{k}$ 10-sec AudioSet~\cite{gemmeke2017audio} examples sampled at $16$~kHz ($T=160$k).
For MixIT training, mixture-of-mixtures examples are synthesized by combining two mixtures at an SNR sampled uniformly from $[-10,10]$~dB. 
We use SI-SNR~\cite{le2019sdr} for the waveform loss $\mathcal{L}$ and an $\ell_2$ embedding loss for $\mathcal{L}_{\mathrm{emb}}$ (using targets from the frozen LAION-CLAP model), with $\lambda=25$. 
Assignments are restricted to proper bipartitions, requiring at least one source estimate per constituent mixture, for $N=4$, these are the $1+3$, $2+2$, and $3+1$ partitions and their permutations.
As a fully supervised separation baseline, we train a Locoformer separator with PIT on the FUSS~\cite{wisdom2021s} training set using isolated-source supervision.

\subsection{Source Separation Evaluation}

We evaluate source separation on the reverberant FUSS~\cite{wisdom2021s} test set. We compare PIT, MixIT, and our proposed method using both Locoformer- and TasNet-based separators. 
We report SI-SNR for single-source mixtures, SI-SNR improvement (SI-SNRi) for multi-source mixtures, as well as under-, equal-, and over-separation rates, corresponding to cases where the number of active estimated sources is lower than, equal to, or greater than the number of reference sources.

\subsection{FUSS Retrieval Benchmark}
\label{sec:benchmark}

In order to evaluate our source-based embeddings we introduce the \textbf{FUSS Retrieval Benchmark}, which re-purposes and labels the FUSS reverberant test set for \emph{zero-shot audio-text retrieval} evaluation.
The FUSS reverberant test set contains 1000 mixtures comprising 2004 unique recordings.
We label each source by joining its FUSS Freesound ID with its sound-class tags in FSD50K~\cite{fonseca2021fsd50k}. 
This labels 1989 recordings, with 15 excluded recordings that are absent from FSD50K, resulting in 999 mixtures. 

Given a test mixture with $K_{\text{src}} \in \{1, 2, 3, 4\}$ active sources, the model produces $N=4$ output source embeddings $\{\bm{e}_i\}_{i=1}^N \in \mathbb{R}^{512}$. We evaluate our source embeddings against three frozen LAION-CLAP baselines:
\begin{itemize}
    \item \textbf{CLAP oracle}: $\operatorname{CLAP}_{\text{audio}}(\bm{x}_k)$, computed from the ground-truth isolated source using the frozen LAION-CLAP audio encoder.
    \item \textbf{CLAP mixture}: $\operatorname{CLAP}_{\text{audio}}(\bm{y})$, computed on the input mixture and used as the embedding of every active source.
    \item \textbf{CLAP post-separation}: $\operatorname{CLAP}_{\text{audio}}(\hat{\bm{x}}_k)$, embedded using the frozen LAION-CLAP audio encoder on the separated waveform.
\end{itemize}
We first embed each sound-class tag in our vocabulary using the frozen text tower $\operatorname{CLAP}_{\mathrm{text}}$.
Because model outputs are unordered, we oracle-align estimated source embeddings to active reference sources. 
For every estimated embedding and reference source pair, we compute the maximum cosine similarity between the source embedding and any tag associated with that reference. 
We then apply Hungarian matching to find the maximum-similarity injective assignment from active references to estimated sources.
After alignment, we rank all vocabulary tags for each source embedding and compute Recall@k as the binary score of whether any of the reference tags appear among the top-k retrieved tags, averaged over all active sources and test mixtures.

\section{Results}
\label{sec:results}

\begin{table*}[t]
\centering
\caption{Evaluation of source separation performance on FUSS.
We compare PIT and MixIT baselines against the proposed method across architectures.
We report SI-SNR for single-source mixtures (1S), SI-SNR improvement (SI-SNRi) for 2-, 3-, 4-, and aggregated 2--4-source mixtures, together with under-, equal-, and over-separation rates. The best MixIT-based result within each model is shown in bold.
}
\label{tab:fuss_results}
\begin{tabular}{llcccccccc}
\toprule
Model & Method & \multicolumn{8}{c}{FUSS (1--4 Source Mixtures)} \\
\cmidrule(lr){3-10}
& & 1S $\uparrow$ & 2Si $\uparrow$ & 3Si $\uparrow$ & 4Si $\uparrow$ & 2--4Si $\uparrow$ & Under $\downarrow$ & Equal $\uparrow$ & Over $\downarrow$ \\
\midrule
Locoformer & PIT & 38.8 & 19.0 & 19.9 & 18.5 & 19.1 & 0.04 & 0.82 & 0.14 \\

\midrule
Locoformer & Audio MixIT & 13.7 & 14.7 & 16.9 & 14.5 & 15.3 & \textbf{0.01} & 0.45 & {0.53} \\
Locoformer & Audio + Latent MixIT & \textbf{21.3} & \textbf{16.1} & \textbf{16.9} & \textbf{14.9} & \textbf{16.0} & 0.08 & \textbf{0.58} & \textbf{0.34} \\
\midrule
TasNet & Audio MixIT & 10.3 & 9.8 & 12.1 & 10.3 & 10.7 & \textbf{0.00} & 0.31 & 0.69 \\
TasNet & Audio + Latent MixIT & \textbf{13.4} & \textbf{12.2} & \textbf{14.4} & \textbf{12.2} & \textbf{12.9} & 0.01 & \textbf{0.36} & \textbf{0.63} \\
\bottomrule
\end{tabular}
\vspace{-0.7cm}
\end{table*}
\begin{table}[t]
\centering
\footnotesize
\setlength{\tabcolsep}{4pt}
\caption{Performance on FUSS Retrieval Benchmark.
The TasNet block forms a controlled ablation of the MixIT objective into audio-only, latent-only, and joint variants, all trained from the same base recipe and teacher.
The best within each section is \underline{underlined}, and best overall is \textbf{bolded}. Supervised baselines (*) are excluded.
}
\label{tab:fuss_semantic_recall}
\vspace{0.2em}
\resizebox{\columnwidth}{!}{
\begin{tabular}{llccc}
\toprule
System & Setup & R@1 $\uparrow$ & R@5 $\uparrow$ & R@10 $\uparrow$ \\
\midrule
\rowcolor{gray!12}\multicolumn{5}{l}{\textbf{CLAP references}} \\
CLAP & Mixture & 17.8 & 37.7 & 51.0 \\
CLAP & GT sources (*) & 46.4 & 73.8 & 82.2 \\
\addlinespace[2pt]
\rowcolor{gray!12}\multicolumn{5}{l}{\textbf{Direct latent retrieval}} \\
- & Latent MixIT & 27.4 & 54.6 & 67.9 \\
TasNet & Audio + Latent MixIT & \underline{32.3} & \underline{60.5} & \underline{\textbf{73.2}} \\
\addlinespace[1pt]
Locoformer & Audio + Latent MixIT & 28.1 & 58.1 & 71.1 \\
\addlinespace[2pt]
\rowcolor{gray!12}\multicolumn{5}{l}{\textbf{CLAP retrieval after separation}} \\
Locoformer & Supervised PIT (*) & 37.2 & 63.7 & 74.4 \\
Locoformer & Audio MixIT & 35.9 & 61.1 & 70.9 \\
TasNet & Audio MixIT & 31.1 & 55.0 & 67.4 \\
Locoformer & Audio + Latent MixIT & \underline{\textbf{36.4}} & \underline{\textbf{61.5}} & \underline{71.8} \\
\addlinespace[1pt]
TasNet & Audio + Latent MixIT & 33.2 & 57.3 & 69.3 \\
\addlinespace[2pt]
\bottomrule
\end{tabular}
}
\end{table}
\begin{table}[t]
\centering
\footnotesize
\setlength{\tabcolsep}{4pt}
\caption{Recall@10 on the FUSS Retrieval Benchmark broken down by source-energy loudness (L1-L4). Loudness rank denotes the ordering of sources within a mixture according to source energy.
The best within each section is \underline{underlined}, and best overall is \textbf{bolded}. Supervised baselines (*) are excluded.}
\label{tab:fuss_semantic_recall_snr_10}
\vspace{0.2em}
\resizebox{\columnwidth}{!}{
\begin{tabular}{llcccc}
\toprule
System & Setup & L1 $\uparrow$ & L2 $\uparrow$ & L3 $\uparrow$ & L4 $\uparrow$ \\
\midrule
\rowcolor{gray!12}\multicolumn{6}{l}{\textbf{CLAP references}} \\
CLAP & Mixture & 67.1 & 45.7 & 34.4 & 31.9 \\
CLAP & GT sources (*) & 83.3 & 80.5 & 82.5 & 81.9 \\
\addlinespace[2pt]
\rowcolor{gray!12}\multicolumn{6}{l}{\textbf{Direct latent retrieval}} \\
- & Latent MixIT & 75.8 & 67.2 & 59.0 & 53.9 \\
TasNet & Audio + Latent MixIT & \underline{78.2} & \underline{\textbf{75.3}} & \underline{\textbf{64.7}} & \underline{\textbf{62.5}} \\
\addlinespace[1pt]
Locoformer & Audio + Latent MixIT & 77.3 & 73.5 & 63.2 & 52.6 \\
\addlinespace[2pt]
\rowcolor{gray!12}\multicolumn{6}{l}{\textbf{CLAP retrieval after separation}} \\
Locoformer & Supervised PIT (*) & 80.0 & 74.0 & 69.9 & 60.3 \\
Locoformer & Audio MixIT & \underline{\textbf{82.1}} & 72.1 & 59.2 & 42.7 \\
Locoformer & Audio + Latent MixIT & 79.7 & \underline{72.9} & \underline{64.1} & \underline{49.6} \\
\addlinespace[1pt]
TasNet & Audio MixIT & 79.8 & 68.0 & 53.6 & 40.1 \\
TasNet & Audio + Latent MixIT & 79.5 & 70.6 & 56.4 & 47.0 \\
\addlinespace[.5pt]
\bottomrule
\end{tabular}
}
\end{table}

In Table \ref{tab:fuss_results}, compared to the MixIT baseline, augmenting the training objective with our proposed latent MixIT one improves or matches separation performance across mixtures with 1 to 4 sources for both TasNet and Locoformer architectures, with uniform gains for TasNet (+2.2 dB on 2–4Si). Gains are largest on single-source inputs (+3.1 dB for TasNet, +7.6 dB for Locoformer), where the proposed approach markedly decreases over-separation (i.e., outputting more active sources than sources in the mixture), while increasing the rate of equal separation. However, this improvement comes with a modest increase in under-separation.
These results suggest that incorporating the latent-space objective provides useful semantic guidance for source decomposition. 
Finally, as expected the fully supervised PIT baseline is the strongest, as it is trained directly with source-level supervision on FUSS and avoids the distribution shift introduced by mixture-of-mixtures training.

Next, in Table~\ref{tab:fuss_semantic_recall} we evaluate the semantic quality of our source embeddings on the FUSS Retrieval Benchmark against frozen CLAP references. CLAP references evaluate embeddings obtained directly from the mixture or ground-truth sources. Direct latent retrieval evaluates the learned source embeddings. For retrieval after separation, the separated waveforms are embedded with CLAP. Text queries use the template ``This is a sound of \{class\}.''
As expected, extracting embeddings from the ground-truth sources performs best, while using the CLAP embedding of the input mixture as a source estimate performs worst.
We highlight that direct latent retrieval with our proposed approach attains the best unsupervised R@10 overall (73.2), exceeding every post-separation baseline while requiring neither the separator nor four passes of the CLAP audio encoder at inference. This suggests that source semantics are largely preserved in the learned latent representations, even though separated waveforms remain preferable at R@1. The two objectives are complementary: latent MixIT alone already far exceeds the mixture baseline (67.9 vs. 51.0 for R@10), adding the audio objective lifts direct latent retrieval by +4.9 R@1, and adding the latent objective lifts post-separation retrieval for both architectures, bringing unsupervised Locoformer (36.4 R@1) within 0.8 points of supervised PIT.

Finally, in Table~\ref{tab:fuss_semantic_recall_snr_10}, we present retrieval performance by source-energy loudness within each mixture. As expected, performance decreases for lower-energy sources across all non-oracle methods, but the degradation is far steeper for separated waveforms (e.g. TasNet: -39.7 points from L1 to L4) than for direct latent retrieval (-15.7). Consequently, the latter overtakes all unsupervised post-separation baselines on the quietest sources, nearly matching supervised PIT.

\section{Conclusion}
\label{sec:conclusion}

In this work, we proposed latent MixIT for source-based semantic representation learning directly from mixtures. We showed that our embeddings support competitive retrieval, while also improving separation when combined with an audio-domain MixIT objective. Future work will explore joint CLAP-style contrastive training of the source representations and text encoder, enabling the latent space to be optimized directly for source-level audio-text alignment instead of relying on a pre-trained embedding model.


\bibliographystyle{IEEEbib}
\bibliography{refs}

\ifarxiv
  \clearpage
  \appendix
  \appendix

\section{FUSS Retrieval Benchmark}
\label{app:fuss}

\subsection{Dataset Construction}

The FUSS Retrieval Benchmark introduced in Sec.~\ref{sec:benchmark} is constructed from the reverberant evaluation split of FUSS (the \texttt{eval/} directory of \texttt{FUSS\_ssdata\_reverb}, exposed as \texttt{test} in TensorFlow Datasets and referred to as the test set in Sec.~\ref{sec:benchmark}).
Each example provides $10$\,s waveforms at $16$\,kHz: the mixture $\bm{y} \in \mathbb{R}^T$ and $K_{\mathrm{src}} \in \{1,2,3,4\}$ reference channels $(\bm{x}_0, \dots, \bm{x}_{K_{\mathrm{src}}-1})$ ordered as \texttt{background0} followed by \texttt{foreground0..2}.
Because FUSS source waveform filenames omit Freesound IDs and \texttt{.jams} annotations are sorted chronologically, we map each event \texttt{annotations[0].data[$k$]} directly to its corresponding channel filename in \texttt{isolated\_events\_audio\_path[$k$]} (under \texttt{annotations[0].sandbox.scaper}) and extract its Freesound ID from the basename stem of \texttt{value.source\_file}.

Each Freesound ID is joined with \texttt{fname} in FSD50K (from files \texttt{dev.csv} and \texttt{eval.csv}), retaining all comma-separated tags from \texttt{labels}.
Across the $1000$ mixtures ($2445$ source instances from $2004$ unique Freesound recordings), excluding the $15$ IDs absent from FSD50K (Tab.~\ref{tab:fuss_excluded}) retains $1989$ unique recordings and $2427$ labeled sources across $999$ mixtures (since a single mixture has only excluded sources).
Mixtures containing excluded IDs are left intact acoustically: the unmodified mixture $\bm{y}$ is fed to the model and excluded sources are omitted only from the mixture's target set $\mathcal{A}$, which contains the reference source indices that constitute valid retrieval targets.
The retrieval tag vocabulary $\mathcal{V}$ comprises all $200$ classes of FSD50K \texttt{vocabulary.csv}.

\subsection{Evaluation Protocol for Retrieval}

Every mixture contains up to $N=4$ reference sources. Therein the $j$-th reference source is associated with a set of tags $Y_j \subset \mathcal{V}$, providing retrieval targets. 
In every mixture the set $\mathcal{A}$ contains all reference source indices that constitute valid targets.
For each mixture, each system produces $N=4$ source embeddings $\{\bm{e}_i\}_{i=1}^N$ whose similarities with vocabulary $\mathcal{V}$ are computed.
For every mixture, Algorithm~\ref{alg:fuss_eval} specifies how source embeddings are aligned with reference sources using Hungarian matching, how references are ranked by loudness, and how retrieval hits are computed.
Final $\mathrm{Recall@}k$ ($k \in \{1,5,10\}$) metrics are computed by micro-averaging over all $2427$ labeled reference-source instances (Tab.~\ref{tab:fuss_semantic_recall}) or within each rank $L_r$ (Tab.~\ref{tab:fuss_semantic_recall_snr_10}) with counts $(N_{L_1}, N_{L_2}, N_{L_3}, N_{L_4}) = (999, 728, 468, 232)$.

\subsection{CLAP Embeddings}

We use the unfused LAION-CLAP variant trained on LAION-Audio-630k, selecting the \texttt{630k-best.pt} checkpoint.
For all tags in $Y_j$ and $c \in \mathcal{V}$, underscores are replaced with spaces and class names are lowercased into \texttt{"This is a sound of \{class\}."} before encoding with $\mathrm{CLAP}_{\mathrm{text}}$ (max token sequence length $77$) and $\ell_2$-normalizing to $\{\bm{t}_c\}_{c \in \mathcal{V}}$.
Audio inputs are resampled from $16$\,kHz to $48$\,kHz before being fed to $\mathrm{CLAP}_{\mathrm{audio}}$.

\subsection{Systems and Baselines}

All systems and baselines return $N=4$ source embeddings $\{\bm{e}_i\}_{i=1}^N$ and are evaluated with the aforementioned evaluation protocol.
Direct latent retrieval uses $\bm{e}_i = \mathrm{Dec}(\bm{s}_i)$ (already unit-normalized); CLAP post-separation uses $\mathrm{CLAP}_{\mathrm{audio}}(\hat{\bm{x}}_i)$; CLAP mixture repeats $\mathrm{CLAP}_{\mathrm{audio}}(\bm{y})$ across all $N=4$ slots; and CLAP oracle (\emph{GT sources}) encodes the $K_{\mathrm{src}}$ reference channels together with $N - K_{\mathrm{src}}$ zero-padded silent channels and applies the same $N \times |\mathcal{A}|$ Hungarian alignment.


\begin{algorithm}[htb]
\caption{Per-Mixture Scoring on FUSS Retrieval Benchmark}
\label{alg:fuss_eval}
\footnotesize
\begin{algorithmic}[1]
\Require Reference sources $\{\bm{x}_j\}_{j=0}^{K_{\mathrm{src}}-1}$ with FSD50K tag sets $\{Y_j\}_{j=0}^{K_{\mathrm{src}}-1}$, vocabulary embeddings $\{\bm{t}_c\}_{c \in \mathcal{V}}$ ($|\mathcal{V}|=200$), source embeddings $\{\bm{e}_i\}_{i=1}^{N}$ ($N=4$).
\State $\mathcal{A} \gets \{j \in \{0,\dots,K_{\mathrm{src}}-1\} : Y_j \neq \emptyset\}$ \Comment{Valid targets}
\If{$\mathcal{A} = \emptyset$} \Return \Comment{Skip mixtures with $0$ labeled sources}
\EndIf
\ForAll{$i \in \{1,\dots,N\},\; j \in \mathcal{A}$}
    \State $S_{i,j} \gets \max_{c \in Y_j} \langle \bm{e}_i, \bm{t}_c \rangle$ \Comment{Multi-label OR similarity}
\EndFor
\State $\pi \gets \arg\max_{\pi: \mathcal{A} \hookrightarrow \{1,\dots,N\}} \sum_{j \in \mathcal{A}} S_{\pi(j), j}$ \Comment{Hungarian match}
\State Rank $j \in \mathcal{A}$ by descending $\|\bm{x}_j\|_2$ to get $r(j) \in \{1,\dots,|\mathcal{A}|\}$
\ForAll{$j \in \mathcal{A}$}
    \State $\mathrm{Top}_k(j) \gets k$ classes $c \in \mathcal{V}$ maximizing $\langle \bm{e}_{\pi(j)}, \bm{t}_c \rangle$
    \State $\mathrm{hit}_k(j) \gets \mathbb{I}\!\left[Y_j \cap \mathrm{Top}_k(j) \neq \emptyset\right]$ \Comment{Pool globally \& into $L_{r(j)}$}
\EndFor
\end{algorithmic}
\end{algorithm}

\begin{table}[htb]
\centering
\caption{Excluded FUSS Freesound IDs missing from FSD50K.}
\label{tab:fuss_excluded}
\vspace{0.2em}
\small
\begin{tabular}{p{0.20\linewidth}p{0.70\linewidth}}
\toprule
Missing (15) &
52293, 131340, 136637, 170811, 170812, 170813, 170814,
170816, 184061, 213181, 320572, 324722, 338196, 366888,
367076 \\
\bottomrule
\end{tabular}
\end{table}

\section{Model Architecture Details}
\label{app:architecture}

Table~\ref{tab:architecture_details} lists architectural configuration and details of every component used in our experiments
(Sec.~4.2--4.3). Table~\ref{tab:architecture_params} presents model trainable parameter count.

\begin{table}[ht]
\centering
\caption{Trainable parameter count per model. 
}
\vspace{0.5em}
\label{tab:architecture_params}
\begin{tabular}{lp{0.45\linewidth}r}
\toprule
\textbf{Model} & \textbf{Components} & \textbf{Params} \\
\midrule
Latent MixIT & $\mathrm{Enc}+\mathrm{Dec}$ & 10.7M \\
TasNet & $\mathrm{Enc}+\mathrm{Dec}+\mathrm{Sep}_{\mathrm{TasNet}}$ & 20.6M \\
Locoformer & $\mathrm{Enc}+\mathrm{Dec}+\mathrm{Sep}_{\mathrm{Locoformer}}$ & 15.1M \\
Locoformer (PIT) & Learned conditioning vectors $+\,\mathrm{Sep}_{\mathrm{Locoformer}}$ & 4.1M \\
\bottomrule
\end{tabular}
\end{table}

\begin{table*}[t]
\centering
\caption{Architecture details and output shapes of each component. Shapes are given for an input mixture of $T$ samples at 16\,kHz, excluding the batch dimension, with $N=4$ slots, slot dimension $C=256$, and CLAP dimension $D=512$. For the PIT baseline, $\mathrm{Enc}$ and $\mathrm{Dec}$ are replaced by $N$ learned 192-dimensional vectors. 
}
\vspace{0.5em}
\label{tab:architecture_details}
\setlength{\tabcolsep}{4pt}
\begin{tabular}{l>{\raggedright\arraybackslash}p{10.3cm}c}
\toprule
Component & Configuration & Output shape \\
\midrule
\rowcolor{gray!15}\multicolumn{3}{l}{\textbf{Embedding model} ($\mathrm{Emb}$, kept frozen)} \\
LAION-CLAP & HTS-AT audio encoder, RoBERTa text encoder & D \\
\rowcolor{gray!15}\multicolumn{3}{l}{\textbf{Slot encoder} ($\mathrm{Enc}: \mathbb{R}^T \to (\mathbb{R}^C)^N$)} \\
Front-end & 64 bin log-mel spectrogram (1024-pt FFT, Hann window, hop 160), per-clip mean/variance normalization & $T_{\mathrm{mel}} \times 64$ \\
Input projection & Linear $64 \to 256$ & $T_{\mathrm{mel}} \times 256$  \\
Backbone & TDCN++: 4 superblocks $\times$ 8 dilated depthwise-conv blocks (dilation $2^0$--$2^7$, hidden width $512$, kernel size $3$) & $T_{\mathrm{mel}} \times 256$ \\
Slot pooling & 
Slot-specific linear projections $256 \to C$, followed by 8-head attention pooling across time using one learned query per slot (attention parameters shared across slots), LayerNorm& $N \times C$  \\
\rowcolor{gray!15}\multicolumn{3}{l}{\textbf{Latent decoder} ($\mathrm{Dec}: (\mathbb{R}^C)^+ \to \mathbb{R}^D$)} \\
Transformer & Learned CLS token; 1 Pre-LN layer (hidden dim $256$, 4 heads, FFN 1024, GELU), no positional embedding, non-CLS tokens cannot attend to CLS & $(1+N) \times C$ \\
Output head & CLS output fed to MLP with GELU activations $256 \to 1024 \to D$, $\ell_2$ normalization & D  \\
\rowcolor{gray!15}\multicolumn{3}{l}{\textbf{Shared separator front-/back-end} ($\mathrm{Sep}: \mathbb{R}^T \times (\mathbb{R}^D)^N \to (\mathbb{R}^T)^N$)} \\
STFT & 512-pt FFT, Hann window, hop 256, $F=257$ bins with real and imaginary parts & $T_{\mathrm{stft}} \times F \times 2$ \\
Masking & Softmax over the $N$ slots, applied as a real mask to the complex mixture STFT, followed by iSTFT & $N \times T$ \\
\rowcolor{gray!15}\multicolumn{3}{l}{\textbf{TasNet separator}} \\
Input & Mixture features tiled over slots; linear $2F \to 256$ & $N \times T_{\mathrm{stft}} \times 256$  \\
Conditioning & FiLM ($\gamma, \beta$) at the input of each superblock, from $\bm{e}_i$ via linear $D \to 2 \times 256$ & $N \times T_{\mathrm{stft}} \times 256$ \\
Core & TDCN++: 4 superblocks $\times$ 8 dilated depthwise-conv blocks (dilation $2^0$--$2^7$, hidden width $512$, kernel size $3$), shared across slots & $N \times T_{\mathrm{stft}} \times 256$ \\
Mask head & Linear $256 \to F$, shared across slots & $T_{\mathrm{stft}} \times F \times N$ \\
\rowcolor{gray!15}\multicolumn{3}{l}{\textbf{Locoformer separator}} \\
Band split & 64 overlapping mel subbands; per-band linear to $192$ & $T_{\mathrm{stft}} \times 64 \times 192$ \\
Initial conditioning & Insert source dimension, source embeddings $\bm{e}_i$ projected with linear $D \to 192$ broadcasted and added over time and bands & $T_{\mathrm{stft}} \times 64 \times N \times 192$ \\
Core & 4 TF-Locoformer blocks (hidden dim $192$), each with: (i) frequency Conv-SwiGLU (hidden 384, depthwise, kernel size $3$) and 4-head RoPE self-attention; (ii) time Conv-SwiGLU and 4-head RoPE self-attention; (iii) 4-head self-attention across slots without positional encoding & $T_{\mathrm{stft}} \times 64 \times N \times 192$ \\
FiLM conditioning & FiLM ($\gamma, \beta$) from $\bm{e}_i$ via linear $D \to 2 \times 192$ in blocks 1 and 3 & $T_{\mathrm{stft}} \times 64 \times N \times 192$ \\
Band merge & Per-band linear, averaged over overlapping bands & $T_{\mathrm{stft}} \times F \times N$ \\
\bottomrule
\end{tabular}
\end{table*}

\fi
\end{document}